\documentclass[preprint,amsmath,amssymb,aps]{revtex4}
\usepackage{url}
\usepackage{indentfirst}
\usepackage{graphicx}
\usepackage{xcolor}
\usepackage{xcolor}

\usepackage{ulem}

\newcommand\dsout{\bgroup\markoverwith{\textcolor{red}{\rule[0.5ex]{1pt}{1pt}}}\ULon}

\begin{document}

\title{Modeling Mito-nuclear Compatibility and its Role in Species Identification}

\author{D\'ebora Princepe}
\affiliation{Instituto de F\'isica `Gleb Wataghin', Universidade Estadual de Campinas - 13083-859, Campinas, SP, Brazil}

\author{Marcus A. M. de Aguiar}
\affiliation{Instituto de F\'isica `Gleb Wataghin', Universidade Estadual de Campinas - 13083-859 Campinas, SP, Brazil}

\markboth%
{Princepe and de Aguiar}
{modeling Mito-nuclear compatibility}

\begin{abstract}
{Mitochondrial genetic material (mtDNA) is widely used for phylogenetic reconstruction and as a barcode for species identification. The utility of mtDNA in these contexts derives from its particular molecular properties, including its high evolutionary rate, uniparental inheritance, and small size. But mtDNA may also play a fundamental role in speciation -- as suggested by recent observations of coevolution with the nuclear DNA, along with the fact that respiration depends on coordination of genes from both sources. Here we study how mito-nuclear interactions affect the accuracy of species identification by mtDNA, as well as the speciation process itself. We simulate the evolution of a population of individuals who carry a recombining nuclear genome and a mitochondrial genome inherited maternally. We compare a null model fitness landscape that lacks any mito-nuclear interaction against a scenario in which interactions influence fitness. Fitness is assigned to individuals according to their mito-nuclear compatibility, which drives the coevolution of the nuclear and mitochondrial genomes. Depending on the model parameters, the population breaks into distinct species and the model output then allows us to analyze the accuracy of mtDNA barcode for species identification. Remarkably, we find that species identification by mtDNA is equally accurate in the presence or absence of mito-nuclear coupling and that the success of the DNA barcode derives mainly from population geographical isolation during speciation. Nevertheless, selection imposed by mito-nuclear compatibility influences the diversification process and leaves signatures in the genetic content and spatial distribution of the populations, in three ways. First, speciation is delayed and the resulting phylogenetic trees are more balanced. Second, clades in the resulting phylogenetic tree correlate more strongly with the spatial distribution of species and clusters of more similar mtDNA's. Third, there is a substantial increase in the intraspecies mtDNA similarity, decreasing the number of alleles substitutions per locus and promoting the conservation of genetic information. We compare the evolutionary patterns observed in our model to empirical data from copepods (\textit{T. californicus}). We find good qualitative agreement in the geographic patterns and the topology of the phylogenetic tree, provided the model includes selection based on mito-nuclear interactions. These results highlight the role of mito-nuclear compatibility in the speciation process and its reconstruction from genetic data.}
{Mito-nuclear coevolution, mtDNA Barcode, Parapatry, Phylogeny}
\end{abstract}

\maketitle

\section{Introduction}

In eukaryotes, cellular respiration occurs in the mitochondrion, an organelle that differs from other cytoplasmic components in having its own genetic material, the mitochondrial DNA (mtDNA). In addition to their metabolic role in the production of energy, mitochondria play a key role in population genetics and evolutionary biology \cite{Avise1987, Ballard2005, Hill2019}. Due to its maternal inheritance,  mtDNA is non-recombinant and thus free of the complications introduced by crossing over. These properties, combined with its high evolutionary rate, make the mtDNA a powerful substrate for inferring the geographic structure of populations (e.g., \cite{Morales2018}), hybridizations \cite{Kearns2018}, and phylogenetic relationships \cite{Pavlova2017}. A standardized segment of the mtDNA, in particular, is used as a genetic marker for species identification in animals \cite {Lane2009}: a region of the cytochrome \textit {c} oxidase I gene, a sub-unit of the respiratory complex \cite{Hebert2003, Forsdyke2017}. DNA barcoding facilitates species delimitation, especially when combined with other taxonomic methods \cite {Tizard2019}, providing accurate identification of vertebrates for both phylogenetically close and distant species. The limitations to the mtDNA-based system occur when typical mtDNA properties are violated \cite{Ladoukakis2017}, such as exceptional cases of recombination and non-exclusive maternal inheritance \cite{Galtier2009}, non-neutral evolution \cite{Dowling2008}, and heteroplasmy \cite{Aryaman2019}. Nevertheless, the high rate of success to distinguish animal species motivates a more fundamental question: why does the barcode work and how does it relate to the nuclear DNA (nDNA) divergences during speciation?

Nuclear and mitochondrial DNA interact during cellular respiration, which requires genes from both sources \cite{Hill2016, Wolff2014}. The mtDNA is a small molecule compared to nDNA, composed of about 15,000-17,000 bases encoding 37 genes, of which 13 genes are involved in  cellular respiration, and 24  responsible for the maintenance of mtDNA itself  \cite{Ballard2004}. The nDNA, on the other hand, contributes about 80 genes to respiration, encoding proteins that are imported into the mitochondrion during the process of oxidative phosphorylation (OXPHOS; \cite{Sunnucks2017}). The mito-nuclear interaction occurs mainly in the electron transport system (ETS) of OXPHOS, where mitochondrial and nuclear-encoded sub-units must be intimately associated along the chain of reactions and, therefore, must be structurally and biochemically compatible \cite{Wolff2014, Hill2017}. To maintain this compatibility, the nuclear and mitochondrial genomes must evolve in a coordinated way, despite their differences in structure, mode of inheritance, recombination, and mutation rates \cite{Bar-Yaacov2012, Sunnucks2017}. Nuclear DNA, with a typically lower mutation rate, must adopt compensatory changes to maintain compatibility with mtDNA, which has a faster rate of molecular evolution \cite{Hill2016}. Mito-nuclear interaction could, therefore, induce the coevolution of proteins encoded by both genomes and facilitate reproductive isolation, as incompatibilities between genes reduce fitness \cite{Sunnucks2017}.

Evidence of mito-nuclear coevolution has been observed in populations that experienced recent reproductive barriers, including positive selection in the nuclear genes that interact with mtDNA \cite{Barreto2018}. Signatures of coevolution have also been found in the process of hybridization \cite {Levin2017}, suggesting that mito-nuclear interaction could  lead to  interpopulation hybrid breakdown \cite{Lima2019}. Although recent results show that mito-nuclear coevolution may be an essential mechanism in speciation, several issues remain unresolved, such as the maintenance of mito-nuclear coadaptation in the presence of gene flow and the relation between mitochondrial metabolism and environmental selection \cite{Hill2019, Hill2019_2}. Studies of mito-nuclear interaction have addressed the speciation problem, employing some recurrent model systems in the literature \cite{Sunnucks2017}. In avian species, for example, DNA barcoding provides highly accurate species identification; it has been suggested that the very concept of avian species could be defined by coadapted sets of nuclear and mitochondrial genes \cite{Hill2017}. The most prominent model animal, though, is the copepod \textit{Tigriopus californicus}, with studies demonstrating the impact of mito-nuclear incompatibilities on individual fitness, eventually causing reproductive barriers \cite{Barreto2018, Lima2019, Burton2012}. Research on interpopulation crosses also suggests the role of mito-nuclear interaction in regulating population structure, combined with restricted gene flow among geographically extended populations \cite{Pereira2016}. These results have intensified interest in the mechanisms underlying the speciation process, from molecular to ecological scales. Thus, understanding the coevolution between DNA and mtDNA and its signatures on population structure and dynamics is a key problem that can benefit from comparison between theoretical models and recent experimental data \cite{Morales2018, Barreto2018}. Moreover, understanding the role of mito-nuclear interaction in the accuracy of species identification by mtDNA can help delimit the scope of reliable barcoding applications.

Here, we investigate the validity of mtDNA-based barcoding in an evolving population, comparing the classification provided by mtDNA with the classification based on nuclear genetic content. We also discuss whether the accuracy of species classification through the mtDNA is affected by the mito-nuclear interaction, i.e., if the mtDNA barcoding is only possible because mito-nuclear coevolution itself is assisting speciation. These analyses allow us to evaluate the effects of mito-nuclear interaction on speciation. 

We use an individual-based model of speciation in which the population is spatially distributed and individuals each have nuclear and mitochondrial genomes. In the first scenario, mating is restricted only by a minimum nuclear genetic similarity, simulating pre-zygotic barriers, and spatial proximity between individuals (parapatry).  All individuals have the same chance to reproduce, provided there are genetically similar mates within their spatial neighborhoods (see \textit{Materials and Methods}). Mating with nearby individuals creates correlations between nuclear genomes and spatial locations;that is, isolation by distance \cite{Wright1943}. We call this scenario the {\it parapatric neutral} model. Depending on model parameters, such as the size of mating neighborhood and the mutation rate, the population can break up into species (radiation) and eventually reach an equilibrium where the (ensemble) average number of species is constant \cite{baptestini_global_2009}. We consider a second scenario where selection is also influenced by  mito-nuclear coupling, considering the compatibility between nucleus and mitochondrion as a component of fitness.  Mito-nuclear interaction is simulated using a much simplified model of the complex process that takes place in the cell: we propose a phenomenological approach to quantify mito-nuclear compatibility based on the lock-and-key principle that guides the interaction between proteins coded by nDNA and mtDNA units.  In the model, each mitochondrial locus is paired with a corresponding locus in the nuclear genome, and we assume that the alleles (bits) in these loci must be compatible to function properly. We also assume that incompatible loci decrease fitness in an additive manner, such that individuals with large number of incompatible loci have small probability of reproducing. A parameter in the fitness function controls the strength of such selection on compatibility. We simulate the evolution of a population formed by initially identical individuals, with maximum mito-nuclear compatibility, and observe how nDNA-mtDNA incompatibilities rise during the initial radiation until a stable number of species is reached. Once equilibrium is reached we finally compare the species identified by the nDNA and the mtDNA genetic content. We analyze signatures of the mito-nuclear interaction in the population's spatial structure and dynamics, storing the number of extant and extinct species, registering their ancestry, and building the corresponding phylogenetic tree. We compare our results with the well-documented empirical phylogeography of populations of \textit{T. californicus}, finding qualitative agreement with patterns obtained from the simulated selective process with mito-nuclear interaction. Finally, we look for signatures of mito-nuclear interaction in the empirical genetic data, calculating allele substitution rates.

\section{Materials and Methods} 

We model nuclear and mitochondrial genomes using a variation of individual-based models for speciation introduced by de Aguiar et al. \cite{baptestini_global_2009}, adapted for populations with separate sexes \cite{baptestini_role_2013} and locus-by-locus recombination \cite{aguiar_speciation_2017}. In the original model, which did not include mtDNA, neutral processes driven by recombination and mutation change allele frequencies in the population. When combined with genetic and spatial restrictions imposed on mating, reproductively isolated species can emerge  \cite{aguiar_speciation_2017}. We adopt the same neutral assumptions here and  introduce mito-nuclear genetic interaction as a non-neutral element, influencing the fitness of individuals. We aim to compare the neutral process, when selection is absent (except for a threshold genetic similarity required for mating compatibility), with the scenario where mito-nuclear compatibility affects fitness, with different levels of selective pressure.

We consider a population of $M$ haploid individuals randomly distributed in a square ($100 \times 100$) or rectangular ($400 \times 25$) lattice with reflecting boundary conditions. The nuclear and mitochondrial genomes are represented by binary strings of sizes $B$ and $B_M$ respectively, $n^i = \{n_1^i, n_2^i, \dots, n_B^i\}$ and $m^i = \{m_1^i, m_2^i, \dots, m_{B_M}^i\}$ for the individual $i$, where each locus $n^i_j$ or $m^i_j$, can assume the allele values 0 or 1. The nuclear genetic distance $\rho^{ij}$ between two individuals $i$ and $j$ is the Hamming distance between the corresponding sequences, and it measures the number of loci bearing different alleles, $\rho^{ij} = \sum_{k=1}^B |n_k^i - n_k^j|$. Similarly, we define the mitochondrial distance $\nu^{ij}$ between two individuals $i$ and $j$ as $\nu^{ij} = \sum_{k=1}^{B_M} |m_k^i - m_k^j|$. We assume that individuals have only one type of mitochondrial configuration and, therefore, we model their genomes with a single copy of mtDNA. In the model, mitochondrial inheritance is strictly maternal and non-recombinant. Sex is assigned randomly with equal probability in the initial population and for offspring (although we discuss biased sex ratios, later). Initially, at time $T=0$, all the genomes' loci are set to zero, i.e., all individuals are genetically identical. We thus emulate the process of the population splitting into multiple species, namely a radiation.

\begin{figure}[th]
\begin{center}
\includegraphics[width=12.5cm]{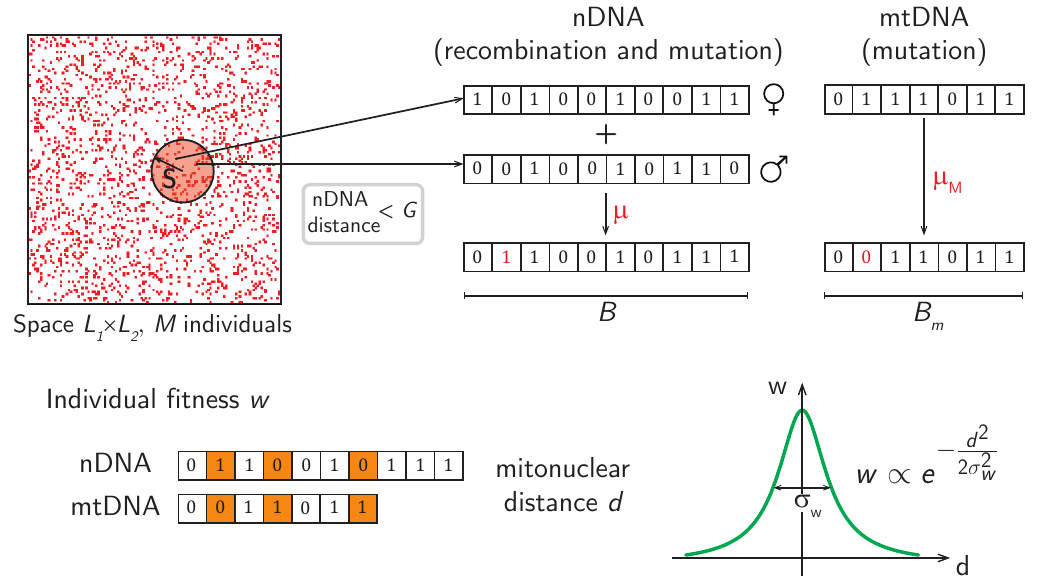}
\end{center}
\caption{Schematic representation of the model dynamics. A population with $M$ individuals (red squares) is distributed on a $L_1 \times L_2$ spatial area, and the mating range is specified by the parameter $S$. The nDNA of compatible parents may differ by up to $G$ bits (loci), and they recombine locus by locus, followed by mutation with probability $\mu$ per locus. The mtDNA is inherited from the mother, with mutation at rate $\mu_M$. The mito-nuclear distance $d$ measures the compatibility between the mtDNA and the corresponding loci of nDNA, assigned to each individual. Here, orange-colored -- eletronic version -- or gray-colored -- printed version -- sites illustrate loci of incompatible alleles between the mtDNA and nDNA genomes. Fitness ($w$) depends on $d$ according to a Gaussian function with width $\sigma_w$, which quantifies the selective strength of the mito-nuclear interaction.}
\label{fig:model}
\end{figure}

Generations do not overlap and are updated as follows: individuals die without reproducing with constant probability $Q$. With probability $1-Q$ individual $i$ chooses as mate any individual $j$ of opposite sex from a circular neighborhood of radius $S$ centered in its spatial location (the \textit{mating range}, see Fig. \ref{fig:model}) whose nuclear genetic distance satisfies $\rho^{ij} \leq G$ \cite{Gavrilets2001}. The mate is chosen uniformly at random among the genetically compatible individuals within the mating range. Speciation only occurs if $S$ is sufficiently small, unless $B$ is very large \cite{aguiar_speciation_2017}. The parameter $G$ represents the \textit{genetic threshold} of reproductive isolation: species are primarily identified as groups of individuals related by genetic distance smaller or equal to $G$ and reproductively isolated from all other groups by this genetic mating restriction. The parameter $G$ was fixed in the simulations as $G=0.05 B$. The offspring inherits, \textit{locus} by \textit{locus}, the nuclear allele of either parent with equal probability and the full mitochondrion from the mother, as illustrated in Figure \ref{fig:model}. After reproduction, the offspring is also subjected to mutation at each locus with probability $\mu$ for the nuclear genome and $\mu_M$ for the mitochondrion. We assume constant mutation rates and independent loci. In order to keep the population size constant, when individual $i$ dies without reproducing (probability $Q$) another individual $k$ in $i$'s mating range is chosen at random to reproduce in its place (more details in section 2 of the {\it Supplementary Information}).

Mito-nuclear interaction and associated fitness are introduced in the model as follows:  (i) the first $B_M$ loci of the nuclear genome interact, locus by locus, with the $B_M$ loci of mtDNA; (ii) the interacting pair is compatible if they have the same allele value ($n^i_k=m^i_k$), and each incompatible pair decreases the fitness of the individual.  Therefore, compatibility is expressed by the similarity between the mtDNA and a portion of the nDNA genomes. Figure \ref{fig:model} illustrates this scheme, where orange shadowed sites -- eletronic version -- or gray shadowed sites -- printed version represent incompatible loci. We define $d^i = \frac{1}{B_M}\sum_{k=1}^{B_M} |n_k^i - m_k^i|$ as the  fraction of sites with different alleles between the two genomes of individual $i$, or the mito-nuclear distance. In our model, $d^i$ is a direct measure of mito-nuclear compatibility in the individual: perfect match ($d^i=0$) indicates full compatibility and maximum fitness, which decreases as the number of mismatching loci increases due to mutations or recombination in the nDNA \cite{Nourmohammad2016}. For completely random sequences, where each allele assumes either value 0 or 1 with equal probability, the expected mito-nuclear distance is $d=0.5$. As in the neutral case, all individuals have a chance of reproducing, which is not the constant $1-Q$, but is proportional to the individual's fitness, defined as $w^i = e^{-(d^i)^2/2 \sigma_w^2}$, normalized within the mating neighborhood. The mate is chosen randomly, with probability proportional to its fitness, among the genetically compatible individuals within the mating range. Thus selection acts on the compatibility between nucleus and mitochondrion, not on their genes separately. When an individual dies without reproducing, another one is chosen from its mating range to reproduce in its place. Again, the choice of this individual is random, but with probability proportional to its fitness. The width $\sigma_w$ represents the strength of selection and is a measure of mito-nuclear distance below which selection becomes intense. Large values of $\sigma_w$ imply a broad fitness function profile, with $w$ weakly dependent on $d$, similar to the flat fitness landscape scenario. When $\sigma_w$ is small, on the other hand, large fitness occurs only for mito-nuclear distance close to zero, and any increase in mito-nuclear incompatibilities lead to a sharp loss of fitness. 

In all simulations, we adopt standard parameter values following previous demonstrations of this speciation model \cite{Costa2018,aguiar_speciation_2017}. The population size is $M=1,300$ and the nuclear genome length is $B=1,500$, with mutation rate $\mu=0.00025$ and genetic similarity threshold $G=75$. We fix the mating range at $S=5$, which is enough to lead to speciation for this genome size. Finally, the probability of dying without leaving offspring is set to  $Q=0.37 \approx e^{-1}$, which is the probability of not selecting one particular individual in $M$ trials with replacement \cite{aguiar_speciation_2017}. With these parameters, we can observe species formation with reasonable computational time, which increases fast with genome size. For the mtDNA we use the fixed values of $B_M=500$ and $\mu_M=3\mu$ so that $\mu_M < 1/B_M$, avoiding the threshold error limit (see comment at the end of this section). To explore regimes of weak and strong selection we varied the fitness function width $\sigma_w$, roughly representing an upper bound to $d$ to which the fitness is still appreciable. From the simulations we found that the range 0.025 to 0.175 gives significant variation from strong selection to close to the neutral case. 

We study the effects and signatures of mito-nuclear coevolution on speciation dynamics and phylogenetic trees. We build the phylogenies recording the time to the most recent common ancestor (MRCAT) between all pairs of individuals at all times \cite{Costa2018} (see  Section 2 of {\it Supplementary Information} for details). We choose two specific metrics for the analysis of the phylogenetic tree topology: the acceleration of diversification, measured by the $\alpha$-value, and tree balance, here represented by the normalized Sackin index $I_n$ -- details on the calculation of these coefficients are described in \cite{Costa2018}. During the population evolution, we use the genetic distance threshold $G$ based on nDNA to classify species as reproductively isolated groups. We also establish an mtDNA-based classification to compare nDNA versus mtDNA barcode accuracy. Because no such criterion regarding reproductive isolation exists for the mitochondrial genome, in our model, we define an artificial mitochondrial genetic threshold $G_M$: individuals are  clustered according to their mitochondrial genetic distances and we identify a `species' (mitotype) as a cluster of individuals whose distance to all others is larger than $G_M$. In the context of graph theory, we map the population into a graph where individuals $i$ and $j$ are connected only if the distance between their mitochondria satisfies $\nu^{ij} \leq G_M$. If $G_M$ is small, there will be few connections and the graph splits into several components (isolated sub-graphs). As $G_M$ increases, more connections are added and the number of components decrease. We search for the value of $G_M$ where the number of components (the mitotypes) is equal to the number of reproductively isolated species. These are represented by components of the network constructed with nuclear distances using parameter $G$ as threshold for connections. Each cluster defines a type (or `species') of mitochondrion, but there is variation within each type since they may differ by up to $G_M$ loci. Therefore, each simulation may produce a different value of $G_M$, depending on the number of species at the final time. We analyze the classification accuracy provided by the mtDNA by checking whether it unambiguously classifies individuals into species in the same way that nDNA similarity does. If two or more species have the same mitochondrion type, or if one species has more than one type of mitochondria, the match is not perfect and classification by mtDNA is considered to have failed. We also analyze how the $G_M$ criterion varies with the strength of the mito-nuclear interaction, providing information about the intraspecies mitochondrial diversity. From now on, we refer to the groups distinguished by the nuclear DNA as species, and we call mitochondrion types or simply mitochondria the classification provided by the mtDNA. All simulations are run in Fortran, and the scripts wrapped in Python are available in the {\it Supplementary Information}. We perform an ensemble of 50 replicate simulations for each set of parameters, unless stated otherwise.

We note that a simpler model of mito-nuclear coevolution can be obtained using the limit of infinitely large genomes, as proposed by Derrida \& Higgs \cite{derrida_evolution_1991}. An analytic expression for the dynamics of the mito-nuclear similarity can be derived in this case if the limit of infinitely many genes is extended to the mitochondrial genome ({\it Supplementary Information}). Fitness is assigned to individuals according to the similarity between their full nuclear and mitochondrial genomes. However, starting from perfectly similar genomes, the stochastic behavior generated by random mutations causes the mito-nuclear similarities to decay exponentially in time for all individuals in the same way. Consequently, the flat fitness landscape is recovered. Differences in fitness arise from the variance of the similarity distribution, but it decays to zero for infinitely large genomes. Thus the finite character of the mitochondria plays a key role in the dynamics of speciation, and we verify that significant variation in individual fitness requires that the mitochondrial genome length satisfy $B_M\simeq  1/\mu_M$, where $\mu_M$ is the mitochondrial mutation rate (more details in Section 1.4 of {\it Supplementary Information}). This derivation resembles the error threshold, that is a critical upper limit for the mutation rate, $\mu_c$, of a finite length sequence compatible with adaptation \cite{Nowak2006}.

\section{Results}

\subsection{Speciation}

\begin{figure}[b]
\begin{center}
\includegraphics[width=12.5cm]{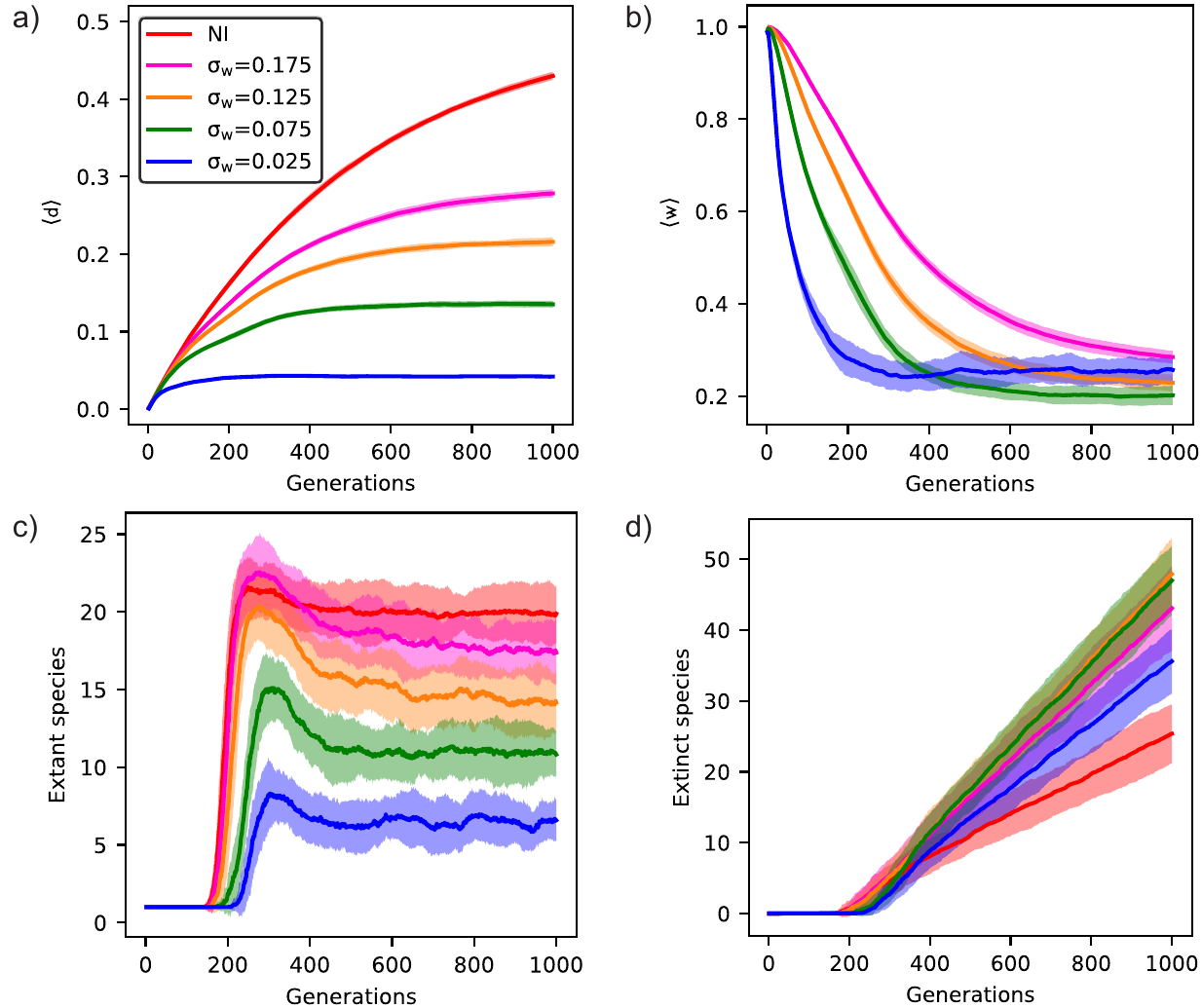}
\end{center}
\caption{Population evolution for different values of the mito-nuclear interaction width $\sigma_w$ (smaller $\sigma_w$ implies stronger selection), and in the absence of any interaction (NI). Results are shown for 50 independent realizations; the shadowed area represents one standard deviation from the ensemble average (solid line). (a) Trajectory of population-averaged mito-nuclear distance. (b) Trajectory of average fitness of the population. (c) The number of extant species. (d) Accumulated number of extinct species.}
\label{fig:spec}
\end{figure}

We simulated the evolution of populations distributed on a square lattice ($100 \times 100$) for $1,000$ generations, about twice the equilibration time (when the number of extant species reaches a plateau). In the non-interacting case (NI), i.e., the neutral scenario, the probability of reproducing was kept constant at $(1-Q)$ for all individuals at all times  (flat fitness landscape). The population-averaged mito-nuclear distance $\left< d \right>$ increased monotonically towards $0.5$, as expected, since the sequences evolve independently (Fig. \ref{fig:spec}a). The introduction of fitness derived from mito-nuclear interaction, on the other hand, progressively lowered the steady-state value of $\left< d \right>$, with the smaller distance occurring for the narrower $\sigma_w$. Thus mito-nuclear interaction preserved mito-nuclear compatibility, as smaller distances represent fewer different alleles between nDNA and mtDNA. Nevertheless, the population-averaged fitness $\left< w \right>$ always decays over time, as shown in Figure \ref{fig:spec}b, since the initial condition corresponds to perfectly compatible genomes and accumulated mutations only add incompatibilities. The stronger the interaction, corresponding to a smaller $\sigma_w$, the faster the decay, because the mutation rate was common in all cases but the limiting distance represented by $\sigma_w$ was smaller and then reached faster. Surprisingly, when the interaction is stronger, corresponding to $\sigma_w=0.025$, equilibrium average fitness was larger, and it also featured larger fluctuations.

We also registered the number of extant species and the accumulated number of extinctions (here taken as species whose population size decreases to zero) as a function of time. Figure \ref{fig:spec}c shows that the mito-nuclear interaction delayed the radiation process and diminished the number of species in equilibrium. It also caused a pronounced overshooting in the number of species before equilibration, which is the decrease of species richness in the later stages of radiation \cite{Meyer2011}. Following an initial transient, the rate of extinctions is constant in time (linear regression of time versus accumulated extinctions $R^2 \geq 0.998$ for all cases). However, the rate of of extinctions  did not follow a monotonic relation with $\sigma_w$ (Fig. \ref{fig:spec}d).  For the neutral case, the rate of extinctions was $0.029$ in units of (time)$^{-1}$, and for the cases with interaction, the values were $0.052$, $0.059$, $0.060$, and $0.045$, from the weaker to stronger selection (i.e., from larger to smaller $\sigma_w$). Therefore, the introduction of fitness increased the extinction rate, which was highest for the intermediate values of $\sigma_w$.

\subsection{Signatures in phylogenetic trees}

\begin{figure}[h]
\begin{center}
    \includegraphics[width=7.6cm]{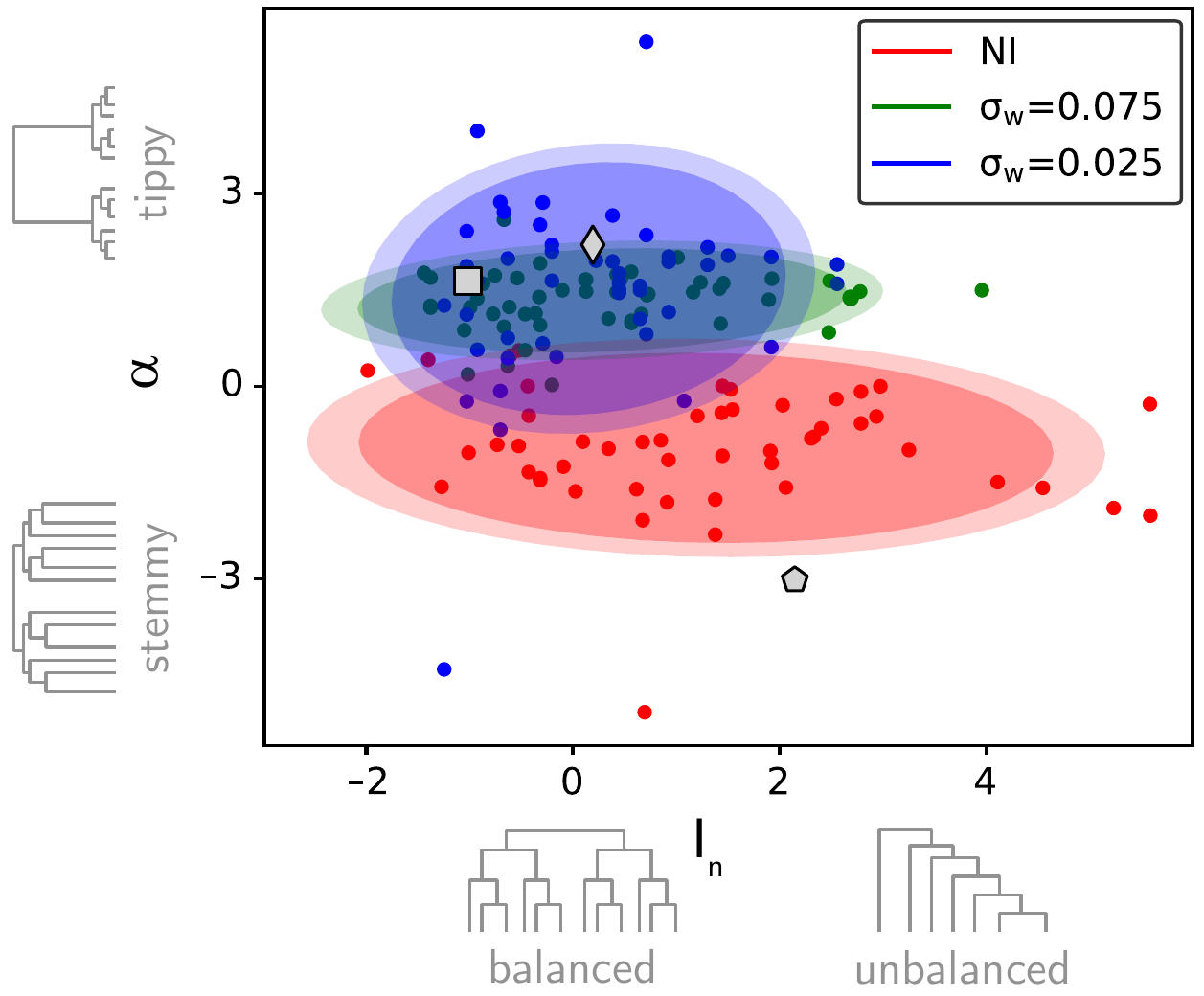}
\end{center}
\caption{Distribution of $\alpha$-values and normalized Sackin index $I_n$ for phylogenies arising from model simulations with or without mito-nuclear interaction (with interaction strength $\sigma_w=0.075$ or $\sigma_w=0.025$), with two spatial dimensions (dots). Ellipses enclose 90\% (darker) and 95\% (lighter) of the simulated realizations for each scenario. The three gray symbols refer to one-dimensional spatial populations: simulations shown in Figure \ref{fig:space} without interaction (pentagon); with interaction, $\sigma_w=0.025$ (square). The diamond indicates $\alpha$ and $I_n$ values for the empirical phylogeny of \textit{Tigriopus californicus} \cite{Barreto2018}. }
\label{fig:tree}
\end{figure}

Different speciation processes produce phylogenies with their own distinct structural properties that can be measured by appropriate metrics. Figure \ref{fig:tree} shows how mito-nuclear interaction influences the acceleration of diversification ($\alpha$-value) and tree balance (normalized Sackin index $I_n$ \cite{Costa2018}) under our model, in the same simulations whose diversity trajectories are shown in Figure \ref{fig:spec}. For each visualization we show results only for the neutral case and for $\sigma_w$ equal to $0.075$ and $0.025$ (results for $\sigma_w=0.0125, 0.175$ overlap considerably with $0.075$). Without selection, the Sackin index varies in a wide range, from balanced ($I_n \approx 0$) to very unbalanced ($I_n > 2$) trees. Diversification was fast at the beginning of the simulation, as the radiation starts, and then slowed down, resulting in negative $\alpha$ values. In contrast, the introduction of a mito-nuclear genetic interaction leads to more balanced trees and accelerated diversification. Also notable is the broader range of $\alpha$-values of the ensemble corresponding to $\sigma_w=0.025$. This behavior is consistent with larger fluctuations observed in the simulations in the case of strong selection.

\subsection{Species identification}

\begin{figure}[t]
\begin{center}
    \includegraphics[width=16.9cm]{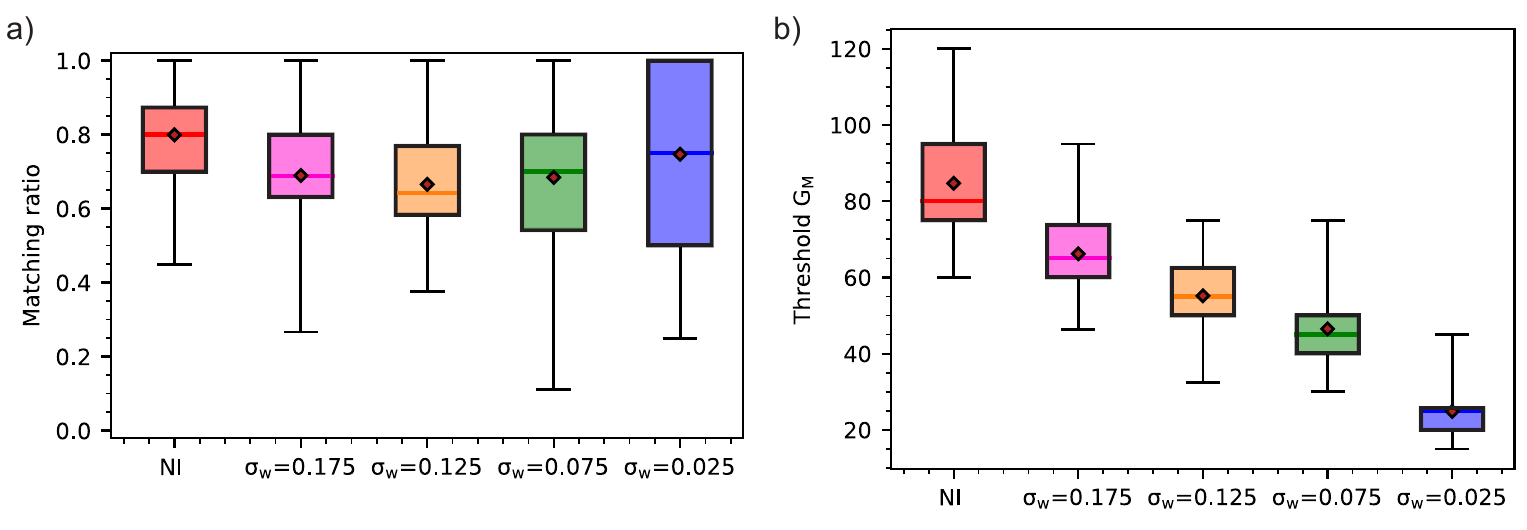}
\end{center}
\caption{Accuracy of the mtDNA barcode for species identification, under our model, in the case of no mito-nuclear interaction (NI) and some interaction ($\sigma_w={0.175,0.125,0.075,0.025}$). (a) Matching ratio of species identified by a unique mitochondrion, as compared to species defined by the nuclear genome. (b) Value of the threshold $G_M$ for each case. Boxplots show the median, lower, and upper quartiles, with whiskers extending to 1.5 times the interquartile range; the diamonds are the average values. Data correspond to 50 simulated replicates of each case.}
\label{fig:barcode}
\end{figure}

We tested whether the success rate of mtDNA species identification is dependent on the mito-nuclear genetic interaction, following the criterion established in \textit{Materials and Methods}. Figure \ref{fig:barcode}a shows the resulting matching statistics. Although we might expect that the interaction would be responsible for the ability to identify species by mtDNA, we find a strong correspondence both in the presence and in the absence of the mito-nuclear coupling. Furthermore, the interaction led to a slight decrease in the average matching percentage, with lower rates of matching for intermediate fitness strength. The stronger interaction also presented a higher variation of the matching ratio. These results imply that, in our model, introducing the mito-nuclear interaction resulted in less precise mtDNA barcoding than the classification in a neutral scenario that has no selection on mito-nuclear compatibility. Therefore the compatibility between species classification using reproductive isolation and the mtDNA barcode is seen more as a consequence of local mating than a result of mito-nuclear coupling. Nevertheless, mito-nuclear coupling has a large effects on the the values of $G_M$, which is a measure of the variance of the distribution of mitochondria within each cluster (Fig. \ref{fig:barcode}b). The average number of different loci in the mtDNA within each species was 80 in the non-interacting case ($G_M=0.16 B_M$, $B_M=500$) and progressively diminished to 20 ($G_M=0.04 B_M$) under strong selection ($\sigma_w=0.025$). Thus  interaction increased the mitochondrial similarity of individuals belonging to the same species, though this did not produce any enhanced ability to identify species through the mtDNA.

\subsection{1-D populations} 

\begin{figure}[b]
\begin{center}
    \includegraphics[width=16.9cm]{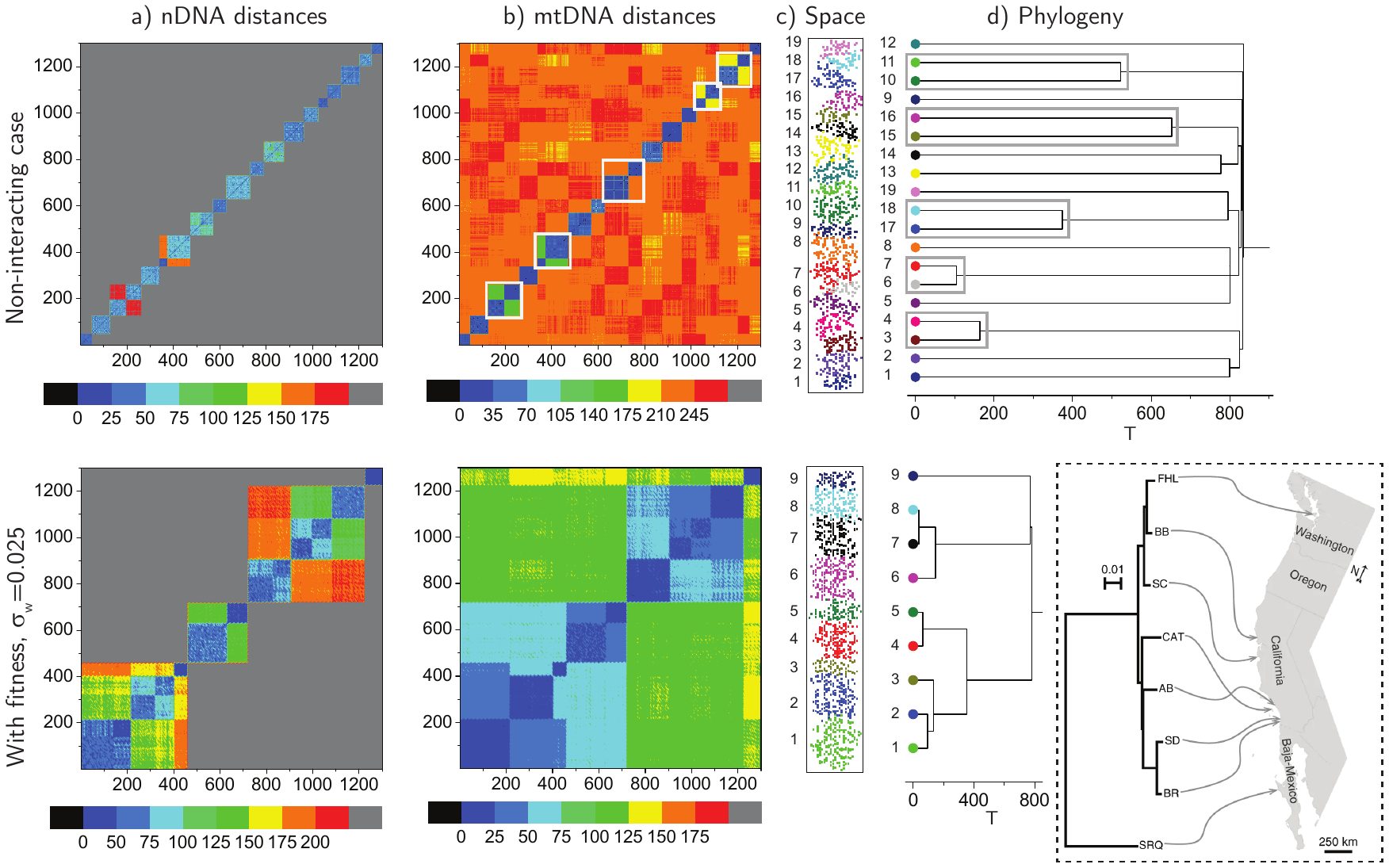}
\end{center}
\caption{Patterns in 1-D populations at $T=1,000$ generations for the non-interacting (NI) scenario (top) and interacting case with $\sigma_w=0.025$ (bottom). Panels (a) and (b) show color maps of genetic distances between pairs of individuals of the population, ordered by species number. (a) nDNA distances -- blocks in the diagonal represent the species. (b) mtDNA distances -- $G_M=135$ (NI) and $G_M=25$ (with interaction). Highlighted speciation events in the NI case correspond to the most recent speciation events, pair of species (3, 4), (6, 7), (10, 11), (15, 16), and (17, 18). (c) Spatial distribution of the population; each color represents a different species. (d) Phylogenies. Bottom right: phylogeny of \textit{Tigriopus californicus} populations, extracted from Figure 1 \cite{Barreto2018} (https://doi.org/10.1038/s41559-018-0588-1) CC BY (http://creativecommons.org/licenses/by/4.0/.).}
\label{fig:space}
\end{figure}

To investigate the influence of geography in speciation and the evolution of mito-nuclear incompatibilities, we modified the original square lattice to an elongated rectangular area, with sides $L_1 \times L_2 = 400 \times 25$. This alternative spatial structure emulates a population spread over a nearly one-dimensional region, as $L_2$ has the same order magnitude as the mating range diameter, $2S=10$. We compared model prediction in two different extremes -- the absence of interaction versus strong selection, $\sigma_w=0.025$. As before, we computed the threshold $G_M$ that distinguished the same number of types of mtDNA as species based on nDNA, producing $G_M=135=0.27B_M$ (non-interacting case, 19 species) and $G_M=25=0.05B_M$ (with interaction, nine species). In both cases, the correspondence between species and mtDNA types was perfect. Figure \ref{fig:space} shows an example of simulation output for a particular realization of each case, after $T=1,000$ generations. The axes of the color maps (Fig. \ref{fig:space}a,b) represent the individuals in the population, and each colored dot at position $(i,j)$ in the map is the genetic distance, nuclear or mitochondrial, between the corresponding individuals. Individuals are conveniently ordered by species, so that the lower  block is species number $1$; the next is species number $2$, and so on. The same ordering was used for the mitochondrial color map. There is clear correspondence between mitochondrial and nuclear distances in both non-interacting (top) and interacting (bottom) cases, with the same arrangement of diagonal blocks, representing the mitochondrion types. However, in the presence of selection based on mito-nuclear interaction the equilibrium number of species is reduced, as are mitochondrial distances within species. As a consequence, mitochondrial genomes were more similar among individuals of the same species with the interaction (Fig. \ref{fig:space}b, bottom). 

The spatial distribution of the simulated population (Fig. \ref{fig:space}c) shows a strong signature of isolation by distance in both selective regimes: more similar species were immediately next to each other. The corresponding phylogenetic trees are shown in Figure  \ref{fig:space}d. We analyzed if there was a correlation between genetic distances and time to the most recent common ancestor. Correlation with the phylogeny was evident in the interacting case: in the genetic color map of mtDNA distances (Fig. \ref{fig:space}b, bottom), we identified three diagonal outer blocks (light blue bounding region, i.e., mtDNA's distance $\leq75$) enclosing the mitochondria (1,2,3,4,5), (6,7,8) and (9). In the phylogeny, they corresponded to clades, with the most ancient speciation event corresponding to the most isolated species in terms of the mitochondrial distance (species 9). This perfect correspondence was not observed, however, without  including mito-nuclear interaction. To illustrate this we highlighted the five most recent branching events in the tree of the non-interacting case and mitochondrial map, noting that the oldest one involves species (15, 16), which remain similar accordingly to the genetic mtDNA color map. Nevertheless, for the second oldest branch, (10, 11), the species have evolved larger mitochondrial distances. Moreover, spatially neighboring species did not always translate into genetic closeness -- species numbered (10, 11 and 12) and (17, 18 and 19), in particular, are swapped in the phylogeny. Therefore, the structure of mtDNA sequences, the spatial distribution of the population, and the phylogeny were much more consistent with each other in the presence of selection on mito-nuclear interactions. The same close correspondence between mtDNA variation, species ranges, and phylogeny is typically observed in empirical systems.

The topological structures of phylogenies showed the same dependence on selection in the one-dimensional spatial setting as we observed in the square lattice. We again find significant differences in the $\alpha$-value and in the Sackin index, indicating that phylogenies produced under strong mito-nuclear interaction are more balanced and feature accelerated diversification compared to the models without selection on mito-nuclear compatibility (Fig.~\ref{fig:tree}). The phylogeny structure for the non-interacting case (pentagon in Fig. \ref{fig:tree}) was roughly consistent with the previous neutral simulations, though it is strictly located outside the ellipse determined by the data from replicates. On the other hand, the scenario with interaction (square in Fig. \ref{fig:tree}) matched the behavior of the previous, two-dimensional simulations with selection, with a large difference from the neutral process.

We compared the outcomes of our model with the phylogeny and spatial distribution of the \textit{Tigriopus californicus} populations, represented in Figure \ref{fig:space} bottom-right (adapted from Figure 1 in \cite{Barreto2018}). Copepods interpopulation show high levels of mitochondrial divergence and hybrid breakdown, presumably in an early stage of reproductive isolation \cite{Lima2019}. These subpopulations are known to have evolved in allopatry \cite{Pereira2016}, with very little gene flow even between geographically close populations \cite{Lima2019}. The model of continuous space used in our simulations is thus intrinsically different from the island structure of these natural populations. Nonetheless, the initial homogeneous population in our parapatric model quickly divides into spatially separated species that evolve, from that point on, as if geographically isolated, with no gene flow between neighbouring species.  This suggests that qualitative comparison between simulation and data is reasonable. Indeed, we find that the empirical tree and spatial distribution of the population shows very similar correlation patterns to those produced by our model assuming strong selection on the mito-nuclear interaction. The location of the populations, related to their phylogenetic proximity, is also qualitatively the same as our model: phylogenetically related populations are also geographically adjacent. We also pursued a more quantitative comparison using metrics of the empirical phylogeny and compared to those in the the 1-D simulations, as shown in the map in Figure \ref{fig:tree} (diamond). The empirical tree metrics follow the predicted behavior assuming a strong mito-nuclear interaction, suggesting that the main aspects of the presumed ongoing speciation are described by our model.

\begin{figure}[h]
\begin{center}
    \includegraphics[width=15.5cm]{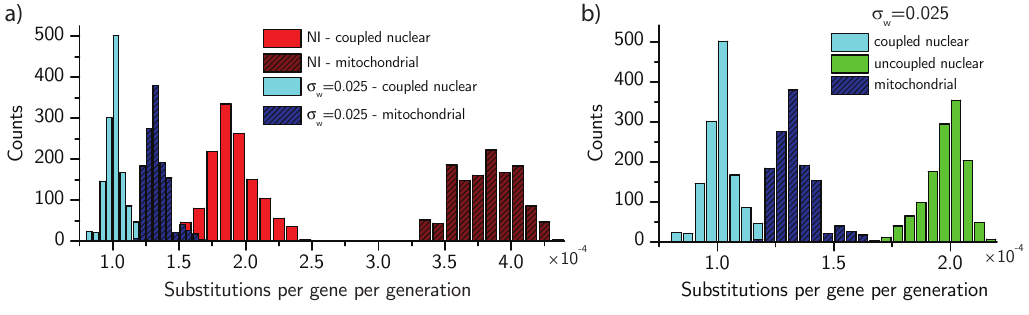}
\end{center}
\caption{Average rate of allele substitutions in the 1-D populations. (a) In the absence of interaction (NI), the substitutions per gene in both the nuclear and mitochondrial genomes correspond to the neutral expectation, $0.78\mu$ and $0.52\mu_M$ per generation. Mito-nuclear coevolution leads to a substantial decrease in substitutions, resulting from strong purifying selection ($\sigma_w=0.025$): substitutions in the interacting nuclear genes drop to around half of the neutral prediction, while the mitochondrial genes retain only 34\% of the predicted number of substitutions under neutrality. (b) Comparison between substitution rates in genes from the nucleus that interact or not with the mtDNA, when there is strong selection on mito-nuclear compatibility.}
\label{fig:mutation}
\end{figure}

Besides macroevolutionary signatures in population geography and phylogenies, we also looked into genetic patterns left in the most recent generation. Mito-nuclear interaction affected the substitution rate of alleles both in the mitochondrial and nuclear genomes. We calculated the number of substitutions, counting the number of genes that changed from the initial configuration by the end of the simulation. Given the mutation rate per locus per generation $\mu$, the probability that a single mutation occurs and is not reversed after $T$ generations is $\binom{T}{1}\mu(1-\mu)^{T-1}$. Since our model allows mutation reversals we must account for the possibility that a mutation occurs and is reversed twice, $\binom{T}{3}\mu^3(1-\mu)^{T-3}$. Thus, $\mu [(1-\mu)^{T-1}+\frac{(T-2)(T-1)}{6}\mu^2(1-\mu)^{T-3}]$ is an estimate for the average number of substitutions per generation (for small $\mu$, the second term is negligible, as well as higher order terms). The expected substitution rates of the nuclear and mitochondrial genomes are then $2.0\times 10^{-4}$ and $3.9\times 10^{-4}$ per generation, respectively. First we analyzed the effect in the interacting genes -- the whole mitochondrial genome and the first $B_M=500$ sites of the nuclear genome (Fig. \ref{fig:mutation}a). Without selection the  substitution rates agreed with the estimated values. The mito-nuclear interaction, however, reduced substantially the rate of allelic substitutions in both genomes: nuclear and mitochondrial sites contained only 51\% and 34\% (respectively) of the number of substitutions that occurred in the non-interacting case. This shows that the interaction produced strong purifying selection on both the genomes, which tended to be more similar to the ancestral condition, therefore promoting the conservation of the genetic information. Nuclear genes that do not interact with mitochondrial genes experienced a substitution rate within the expected values, as shown in Figure \ref{fig:mutation}b. 

\section{Discussion}

Nuclear genes that participate in  respiration exhibit different signature of selection as compared to the rest of the genome \cite{Barreto2018}, which provides evidence of mito-nuclear coevolution. Genomic mapping of isolated populations has shown such differences, which reflects a broad correlation between the requirement for mito-nuclear compatibility and reproductive isolation \cite{Pereira2016, Lima2019}. However, it is not clear if mito-nuclear coevolution is a cause or effect of the speciation process, and if this is the reason for the success of the mtDNA barcode identification of species. We have proposed a model for mito-nuclear interaction that allows us to investigate its influence in parapatric speciation and in the accuracy of DNA barcodes for species classification. Comparing our model behavior with or without selection on mito-nuclear compatibility we have characterized the patterns that are caused by the mito-nuclear interaction and the resulting geographical mode of speciation. 

We found that the presence of selection induced by the mito-nuclear coupling results in fewer species, in equilibrium, and higher extinction rates. Also, we observed a more pronounced overshoot in species number before equilibration was reached. Such effect has been previously considered a signature of adaptive radiation \cite{Meyer2011}, although this inference is not unambiguous as it might also arise in neutral radiations if genome size is very large \cite{Costa2018}. The interaction in our model delayed the radiation process, with a later acceleration of diversification. This effect can be the result of the initial fast decay of fitness, which was faster for stronger interactions (Fig. \ref{fig:spec}b). Later stabilization of the mito-nuclear distance (Fig. \ref{fig:spec}a) led to subsequent accelerated speciation rate. By contrast, the mito-nuclear distance increased steadily in the non-interacting case, the speed of diversification had the opposite behavior: speciation was fast in the beginning and then slowed down. These dynamics are reflected by the tippy phylogenies (measured by the $\alpha$-value) observed under our model with selection when compared to the neutral model. The interaction also produced more balanced trees, compared to neutrality.

Differences in the topological properties of phylogenetic trees allow comparison with empirical speciation events and even the inference of conditions driving evolution \cite{Morlon2014,Gascuel2015,Costa2018}. In our model, apart from selection imposed by the mito-nuclear genetic interaction, the spatial distribution of the population also affects the pattern of radiation. To isolate the effects of selection from those of geography, we considered a sympatric neutral scenario extending the Derrida-Higgs infinite genes model ({\it Supplementary Information}). Local mating can itself delay the radiation process and lead to larger number of species than random  mating (Fig. S1(a)). Nevertheless, such a delay was enhanced when the mito-nuclear interaction was introduced, and the number of species was lowered further. Interestingly, the neutral sympatric speciation process with infinite genomes and the parapatry with selection produced very similar trees (Fig. S1(b)) -- indeed, adaptive and neutral radiations sometimes exhibit similar patterns \cite{Costa2018}. Differentiation between the two processes requires inspection of other aspects of the dynamics, such as the number of extant species, delay of the radiation, extinction rate, or directly in the spatio-genetic patterns.

Our model suggests that barcode identification of species will work regardless of the presence of interaction. Its success in our model is attributed to the geographical mode of speciation. In empirical systems, strong mito-nuclear selection in birds has been used to explain why mtDNA identification works so much better for avian species than for mammals or plants \cite{Hill2017}. Our results contradict this explanation, as we did not find significant differences in accuracy between the cases with and without the interaction. Compared with neutral sympatric speciation (Fig. S1(c,d)), there was an enhancement in the unambiguous species identification when local mating was introduced. In other words, the spatial structure of the population promoted the isolation and uniformity of mtDNA within a species, facilitating its use in species identification. 

The barcode criterion we adopted was based on matching the number of species with mitochondrial clusters. Genetic patterns of the populations support this choice: while both uncoupled and coupled cases produce similar dispersion of nuclear genetic distances within species, centered on the genetic threshold (Fig. S4(a,b)), the mitochondrial distances displayed  distinct distributions, both in the envelope profile and in the range. Strong interaction led to more similar mitochondria among conspecifics (Fig. S4(c,d)). These results suggest that the shape of mitochondrial distance distribution provide a clear signature of selection on mito-nuclear compatibility. Indeed, the separation between mitochondria are sharper and more uniform within species, in our model, in the presence of such an interaction. We also compared the inter-specific mitochondrial distances produced by our model to data from \textit{T. californicus} populations (see Tables S1 and S2 in the {\it Supplementary Information}). The divergence ratios were comparable, and the average values were very close: 19.6\% across all the copepod populations and 16.8\% in the simulation with interaction. Without the interaction, divergences between different species are in average of 44.6\%. This unrealistically large number suggests that mitonuclear interactions are always present in some degree.  

The effects of selection on the number of extant and extinct species, as well as the overshooting effect, suggest that hybridizations occurred in model, and that their rate were affected by the interaction. Hybridizations occur when recently formed species merge together due to a mutation that decreases the distance between reproductively isolated individuals. The occurrence of hybridizations is pertinent to the maintenance of mito-nuclear compatibility in introgressions \cite{Hill2019_2}. Hybridization events occur more readily in the presence of mito-nuclear interaction, as shown in {\it Supplementary Information} (Fig. S3). When compared with the number of living species, both extinction (Fig. S3(a)) and hybridization (Fig. S3(b)) rates monotonically increase with the strength of selection.  Furthermore, the spatial distribution of the population itself was also important for species survival, as 1-D populations presented fewer extinctions and hybridizations, especially in the scenario with selection on mito-nuclear compatibility (Fig. S3(c,d)). The more neighboring species, as in the 2-D configuration, the higher is the probability that a species will either go extinct or hybridize.  These finding suggest future research directions on the role of mito-nuclear coevolution in hybridization. An interesting empirical question, for example, is whether mtDNA barcoding performs poorly in recently merged species.

The introduction of fitness also led to stronger correlation between the mitochondrial genetic content and the spatial distribution of species and phylogenies. We observed the correspondence between clades and clusters of similar mtDNAs, concluding that the mitochondrial similarities reflected better the history of the species in the model with selection. We compared the phylogeny and spatial distribution of our model to those observed in \textit{T. californicus} populations, a species known for strong signatures of mito-nuclear coevolution \cite{Lima2019,Sunnucks2017} and for exhibiting mitochondrial divergences between natural populations \cite{Barreto2018}. Although there is evidence that these populations evolved islands of strict geographic isolation \cite{Pereira2016}, our model with local mating (and, therefore, restricted gene flow) also results in spatially separated species. Post-zygotic mechanisms also play important roles in the emergence of reproductive barriers in copepods \cite{Foley2013}, since mito-nuclear incompatibilities may cause inviability of hybrid offspring \cite{Burton2012, Lima2019}. This feature is also present in our simulations, as we assign fitness according to mito-nuclear compatibility. Genetic similarity and spatial proximity both influence reproduction as pre-zygotic mechanisms, in our model; and mito-nuclear compatibility affects the individual's chance of reproduction, acting as a post-zygotic barrier, as it also affects the fitness of offspring. Surely our model does not capture all the biological aspects of copepods, though. We did not take into account, for example, the mechanisms of sex determination \cite{Alexander2016} that may lead to sex-biased populations \cite{Voordouw2002,Alexander2016}. A lower number of females would decrease mitochondrial variability and moderate the effects of selection (due to a smaller effective population of mitochondria), but we expect no qualitative changes in the  characteristics of the populations.  To confirm this hypothesis we ran simulations with female to male ratio of 40\% and observed similar number of extant species and phylogeny structures. Finally, our model predicts conservation of the genetic material due to mito-nuclear interaction. The introduction of fitness resulted in more similar mitochondria even between different species, which can be interpreted as a signature of conservation or purifying selection. A direct measure of this was provided by number of substitutions per locus, which decreased considerably with the interaction and coevolution of the nuclear and mitochondrial DNAs. This relates to the trade-off between purifying selection on mtDNA and mtDNA-driven positive selection on nDNA \cite{Barreto2018}. Future studies might provide in-depth investigation on the role of the mitochondrial mutation rate and its subsequent effects on the nuclear and mitochondrial substitution rates. Our model provides a platform to examine these issues and the role of the mtDNA mutational rate in the evolution of species \cite{Nabholz2009}.

Our results help to elucidate the role of mito-nuclear interactions in the evolution of spatially distributed populations, including the speciation process. Compared to the complexities of protein-protein interactions, our model is a minimalist scheme of mito-nuclear interaction. The model neglects epistatic effects and, more importantly, considers a one-to-one correspondence between interacting loci the mitochondrial and nuclear genomes. Both of these simplifications are unrealistic. Although it is possible to construct more complex models where many nuclear loci interact with one or more mitochondrial loci, this would introduce complications of interpretation that we wish to avoid. Previous mathematical models that take into account physiological processes described some aspects of the mito-nuclear coevolution \cite{DeBivort2007, Connallon2018}.  However, it is not clear how to scale these effects to the population level without loss of details. In this context, our model, though simplistic, provides insight into how the interaction at the gene level may affect population evolution. Similar models have been used before to describe genetic interactions in the context of coevolving competing entities \cite{Nourmohammad2016}, showing that emerging properties of the system can be revealed by such a schematic description. A further extension of our model would include the possibility of speciation driven exclusively by the mito-nuclear interaction, removing the genetic similarity restriction for mating and the mating range, which are mandatory elements in the neutral model to induce speciation. Selection imposed on mito-nuclear compatibility alone might reproduce the coadaptation scheme present in nature and possibly lead to the emergence of new species. Without the genetic compatibility threshold, the species barrier also has to be considered, providing an opportunity to test the theory of mito-nuclear compatibility species \cite{Hill2017}. 

\section{Supplementary Material}
Data available from the Dryad Digital Repository: https://doi.org/10.5061/dryad.2ngf1vhk7.

\section{Funding}
This work was supported by the S\~ao Paulo Research Foundation (FAPESP), grants \#2016/01343-7
(D.P. and M.A.M.A, ICTP-SAIFR), \#2018/11187-8 (D.P.) and \#2019/20271-5 (M.A.M.A.), by the Coordena\c c\~ao de Aperfei\c coamento de Pessoal de N\'ivel Superior - Brasil (CAPES) - Finance Code 001, and by CNPq, grant \#302049/2015-0 (M.A.M.A.).

\section{Acknowledgements}
We are very grateful to Prof. Felipe Barreto for the support on the dataset of the \textit{T. californicus} interpopulation tree. We also thank Dr. Flavia M. D. Marquitti for fruitful discussions. Special thanks to Dr. Joshua Plotkin for the many suggestions and careful reading of our manuscript.

\end{document}